\documentclass[acmsmall]{acmart}
\usepackage{subfig}
\usepackage{amsmath}
\usepackage{xcolor}
\usepackage{multirow}
\usepackage{graphicx}
\usepackage{xspace}
\usepackage{enumitem}
\usepackage{amsmath}
\usepackage{booktabs}
\usepackage{float}

\AtBeginDocument{%
  }

\setcopyright{acmlicensed}
\copyrightyear{2024}
\acmYear{2024}
\acmDOI{XXXXXXX.XXXXXXX}

 \acmJournal{PACMHCI}
\acmVolume{00}
\acmNumber{0}
\acmArticle{000}
\acmMonth{10}

\begin{document}

\title{Effect of AI Performance, Perceived Risk, and Trust on Human Dependence in Deepfake Detection AI System}

\author{Yingfan Zhou}
\affiliation{%
  \institution{The Pennsylvania State University}
  \country{USA}
}
\author{Ester Chen}
\affiliation{%
  \institution{Rochester Institute of Technology}
  \country{USA}
}
\author{Manasa Pisipati}
\affiliation{%
  \institution{The Pennsylvania State University}
  \country{USA}
}

\author{Aiping Xiong}
\affiliation{%
  \institution{The Pennsylvania State University}
  \country{USA}
}
\author{Sarah Rajtmajer}
\affiliation{%
  \institution{The Pennsylvania State University}
  \country{USA}
}


\begin{abstract}
Synthetic images, audio, and video can now be generated and edited by Artificial Intelligence (AI). In particular, the malicious use of synthetic data has raised concerns about potential harms to cybersecurity, personal privacy, and public trust. Although AI-based detection tools exist to help identify synthetic content, their limitations often lead to user mistrust and confusion between real and fake content. This study examines the role of AI performance in influencing human trust and decision making in synthetic data identification. Through an online human subject experiment involving 400 participants, we examined how varying AI performance impacts human trust and dependence on AI in deepfake detection. Our findings indicate how participants calibrate their dependence on AI based on their perceived risk and the prediction results provided by AI. These insights contribute to the development of transparent and explainable AI systems that better support everyday users in mitigating the harms of synthetic media.

\end{abstract}



\maketitle

\section{Introduction}
Artificial intelligence (AI) technologies now support the generation and editing of synthetic audio, images, or videos based on textual descriptions \cite{epstein2023art}. As generative AI tools have become popular and abundant in recent years, even users without technical background or artistic skill can easily merge, combine, replace, overlay, and generate images and videos to produce synthetic multimodal content \cite{lee2024deepfakes}. Synthetic images improve creativity, reduce costs, and improve efficiency in content creation. 

However, there is growing concern about the harms induced by AI-generated audio and visual content \cite{epstein2023art}. The term 'deepfake', a combination of the words 'deep learning' and 'fake,' is used to describe maliciously used synthetic media content that fabricates or alters someone saying or doing something\cite{westerlund2019emergence}. Deepfake represents an increasing threat to public safety in national security, law enforcement, finance, and society. The cybersecurity challenges induced by deepfakes include, but are not limited to, terrorist propaganda, blackmail, fake pornography, cyber kidnapping scams, cyberbullying, and market manipulation \cite{deepfakeReport, mink2022deepphish}. %Moreover, with the rapid advancement of generative AI technology, synthetic images are becoming increasingly complex to differentiate from real ones. Human struggle to distinguish between artificially generated images and authentic ones. The blurring of the line between real and fake content undermines trust, making it difficult for individuals to discern what is true, which can result in harmful consequences like uninformed decision-making. 

Previous studies have shown that humans struggle to distinguish between deepfake images and real ones. In response to the spread of synthetic images on social media and the potential harm caused by deepfakes, the field of deepfake detection has gained significant attention, driving the development of various AI deepfake detection techniques. For example, social media platforms, such as Instagram, have now applied the deepfake detection AI system to create "Created By AI" warning labels. Warning labels inform users that the content is identified as synthetic content by the authors or platforms\cite{suciu_created_nodate}, which aims to provide transparency and additional contextual information to help readers differentiate real information. The platform noted that the design of the warning label is expected to decrease the risk of deceiving the public about an important issue\cite{noauthor_our_2024}. In cases like these, AI systems provide information while humans make the final decision. This paradigm is applied to various fields such as healthcare, finance, criminal justice, etc. HCI (Human-Computer Interaction) researchers refer to it as AI-assisted decision making.

Previous work has discussed how algorithmic errors can lead to mistrust and confusion among system users, such as misremoving legitimate content in content moderation \cite{paudel2024enabling} or incorrect responses from voice assistants \cite{baughan2023mixed}.  Today, AI systems are often still limited in their performance and can exhibit some biases.  Deepfake detection techniques cannot always accurately differentiate between real and deepfake content. Errors in AI detection algorithms can further contribute to the confusion surrounding the deepfake content. Therefore, it is essential to explore whether and how error warning labels from AI systems influence individual decision making in the context of deepfake detection.

Various studies explored the mechanisms to improve AI-assisted decision making performance and noted that one of the key components is human dependence on AI \cite{zhang2022you,cao2022understanding,salimzadeh2024dealing}. Humans should appropriately rely on the information provided by AI, neither fully relying on it nor completely disregarding it. Researchers have explored the different factors that impact human dependence on AI, including human factors (e.g., personal factors \cite{ma2023should}, trust in the AI system \cite{buccinca2021trust}), transparency and performance of the AI system \cite{lu2021human}, and task factors (e.g., difficulty, uncertainty)\cite{salimzadeh2024dealing,salimzadeh2024dealing}. However, compared to other AI-assisted decision making tasks, the deepfake detection task is somewhat special. First, humans typically lack the ability to perform this task independently as they cannot reliably identify whether an image is a deepfake or assess the accuracy of their judgment. Second, deepfakes may lead to harm or loss. Humans' perceived risk regarding potential loss caused by deepfakes may impact their decision making with the support of AI systems. We can draw an analogy to the operator scenario in human-automation interactions, where the operator often relies on signals provided by the automated machine to perform complex tasks. However, an AI system is different from a machine, as the decision making process of an AI system cannot be clearly explained in steps. Whether theories in the prior paradigm of human-automation interaction can be adopted in a new generation as human-AI interaction is unclear.

In this study, we build on the established theory in human-automation interaction to evaluate the factors that impact users' decision making processes with the support of AI deepfake detection algorithms. We aim to explore how warning labels provided by AI systems assist human decision making. The following research questions drive the work:

\begin{itemize}
\item \textbf{RQ1:} How does the performance of AI deepfake detection algorithms impact human dependence, operationalized through compliance and reliance, on these algorithms?

\item \textbf{RQ2:} How does the performance of AI deepfake detection algorithms impact human trust towards these algorithms?

\item \textbf{RQ3:} How does human trust mediate human dependence on AI deepfake detection algorithms?
\end{itemize}

To address these research questions, we %recruited 400 participants on the Prolific platform and conducted a user study. We 
designed and ran an online experiment in which we asked 400 participants to differentiate between real and synthetic images with the support of outputs from notional AI deepfake detection algorithms with varying levels of performance. %under AI-provided information with varying AI performance levels.
Through moderated mediation analysis, we explored the relationships between AI system performance, human perceived risk of synthetic images, human trust in AI, and whether humans would adopt the AI's judgments, particularly in a transparent setting.

Our study contributes to the understanding of human trust and dependence in AI. In situations where humans lack the ability to distinguish what is real from what is artificially generated, they adjust their decisions with AI-provided information based on the AI system’s performance. We found that when humans learn that an AI misclassifies real images as synthetic, they are more likely to believe that the images classified as real by the AI are indeed real. Our findings provide insights on how to improve the design of future AI systems to better support individuals by reducing the harm caused by the malicious use of synthetic images.

\section{Related Work}
Our study is situated in two areas: deepfakes and human-AI decision making.

\subsection{Generative AI and deepfake}

Generative AI is a type of AI algorithm that can create new content, including text, images, audio, and video \cite{epstein2023art}. "Deepfake," a combination of "deep learning" and "fake," is used to describe synthetic media content \cite{westerlund2019emergence}.

Deepfakes present challenges and threats to public discourse and democratic processes, leading to a growing body of research focused on mitigating potential harms \cite{lee2024deepfakes}. Researchers in the fields of machine learning and computer vision have begun to explore new technology to detect synthetic content \cite{li2023unganable, blue2022you}.

At the same time, HCI researchers focus on understanding people's perception of deepfakes. Past field studies have been conducted to explore how humans from different groups perceive deepfakes \cite{han2024uncovering}. For example, through a mixed-method analysis of conversations in Reddit communities \cite{gamage2022deepfakes} and interviews in low-resource communities \cite{shahid2022matches}, researchers found that users were generally unaware of the potential harm of AI-generated content. Although some express concern about deepfake videos, overall awareness of this issue remains low. They proposed that future platforms and educational initiatives should communicate the risks associated with this type of content to users. Past studies have also examined factors influencing individual abilities to identify synthetic content, such as biases, stereotypes, prior training, and social context \cite{mink2022deepphish, groh2022deepfake, nightingale2022ai}. Moreover, past researchers have explored how existing technologies can support people in identifying deepfake images and have identified current sociotechnical gaps. To enhance human ability, researchers have compared human and machine learning models to examine which types of images are harder for people to recognize \cite{tahir2021seeing}. The researchers found differences in focal points between humans and detection algorithms when identifying images, leading them to propose contextualized education and training to improve detection skills \cite{zhou2023synthetic,tahir2021seeing}.  

Previous HCI literature has emphasized the importance of addressing gaps in current laws and policies, platform designs, improving awareness through education, and the technical support of detection models \cite{lee2024deepfakes}. From the individual user’s perspective, it remains unclear whether they can effectively mitigate risks with the implementation of detection technology. Therefore, our study aims to explore the factors that impact individuals to identify deepfakes with technological support.

\subsection{Human-AI decision making}
Over the past few decades, due to the complexity and ubiquity of automation, operators often assume the role of monitors, relying on system-provided signals to make decisions. There are two types of dependence: compliance and reliance. Compliance refers to operators responding to a signal of automation. Reliance refers to operators refraining from reacting when the system is silent or when it indicates normal operation \cite{meyer2001effects}. However, systems are not always accurate, and excessive or insufficient dependence can lead to misuse or disuse. To optimize performance, researchers in the interaction between human and automatic processes explored the mechanisms of dependence on automated systems \cite{rice2009examining,meyer2001effects}. They found that human dependence on automation is impacted by human-related factors (e.g., trust, emotion processes, individual skill levels), automation-related factors (e.g., error types, system accuracy, additional information provided) and task-related factors (e.g., time pressure, risk, workload)\cite{meyer2004conceptual}.

In recent years, AI has been applied to support human decision making in areas such as law, healthcare, and business \cite{lai2023towards}. There have been extensive concerns about AI-generated decisions, as AI lacks explainability and transparency \cite{lu2021human,dwivedi2023explainable}. Furthermore, the uncertainty and complexity of AI have sparked HCI research into human-AI collaboration, particularly in the context of human-AI decision making. It refers to settings where machine learning models assist users in making final judgments or decisions, forming part of a reasonable collaboration between humans and AI \cite{buccinca2021trust}. For example, in the context of deepfakes, detection models are used to moderate online content and alert users to reconsider the authenticity of the content. Previous researchers have studied how people accept and understand AI decisions, while current research focuses on identifying which forms of AI tools can help humans surpass human and AI capabilities, achieving complementary performance between humans and machines \cite{liu2021understanding, zhang2022you,chiang2023two,vaccaro_when_2024}. This complementarity is reflected in a dynamic similar to human-automation interaction, where users rely on AI when it is accurate but can revert to their own judgment when the AI makes errors. HCI researchers explored human-AI complementarity through the lens of human dependence on AI \cite{schemmer2022should,buccinca2021trust, chen2023understanding}. Previous work has highlighted trust as a critical factor which impacts both human reliance and compliance on AI \cite{rechkemmer2022confidence,buccinca2021trust}.Trust is referred to “the trustor’s willingness to assume risk by delegating responsibility to the trustee” \cite{mayer1995integrative}. When trust is miscalibrated, it can hinder effective collaboration \cite{lee1994trust, lee2004trust}. Overtrust may lead to blind acceptance of incorrect AI judgments, while distrust can cause rejection of  suggestions, undermining AI’s intended role as a decision support tool. Further research has shown that human trust in AI systems does not directly determine dependence on them \cite{vereschak2024trust,kirlik1993modeling}. Trust functions as an attitude that impacts, but does not fully determine users’ behavioral responses, including reliance and compliance. This points to a more nuanced relationship between trust and human dependence on AI \cite{salimzadeh2024dealing,kirlik1993modeling, ma2023should}. Emerging study pointed out whether the human chooses to follow the AI’s decision or rely on own judgment can be impacted by human-related factors, such as human intuition \cite{chen2023understanding}, self-confidence \cite{rechkemmer2022confidence}, trust in AI \cite{rechkemmer2022confidence,buccinca2021trust}, and human likelihood of correctness \cite{ma2023should}. It is also influenced by attributes of the AI itself, including system transparency \cite{chen2023understanding}, perceived AI performance \cite{lu2021human}, and explainability \cite{dwivedi2023explainable}. Task characteristics are also studied, such as uncertainty, and the difficulty\cite{salimzadeh2024doubt,salimzadeh2024dealing}. Drawing on these findings, HCI scholars have proposed ways to avoid cognitive overload and ineffective reliance strategies. The studies aim to enhance individuals' understanding of AI, including its performance\cite{bansal2019beyond,cabrera2023improving}, explainability\cite{dwivedi2023explainable,chen2023understanding}, and uncertainty\cite{wischnewski2023measuring,zhao2023evaluating}.

The task scenarios explored in the past studies explain how people have some capacity to assess and compare their own decision making performance. However, specifically in the context of deepfake tasks, according to previous literature, people lack the ability to judge their task performance. Especially in real-world contexts, individuals cannot compare against the actual truth or real content to understand either the systems or their own performance. Moreover, the confusion between real and deepfake images might induce potential harm. How people rely on AI systems when they lack the ability in such a context remains unclear.

\subsection{Research Question and Hypothesis}

In deepfake detection tasks, the AI system typically serves as a warning provider. In this scenario, an individual user’s task capability and understanding of the system are similar to that of an operator facing a task in human-automation interaction. For example, in human-automation interaction, due to the complexity of automated systems and the duration and difficulty of tasks, the operator plays the role of a monitor, responding to signals from automation \cite{meyer2001effects}. Similarly, in deepfake detection tasks, individual users lack the ability to identify deepfake images themselves, and untrained users are not familiar with the steps by which the AI system generates a deepfake alert. Upon receiving a warning, the user needs to decide whether to rely on the AI's predicted results or warnings.

Past studies in human-automation interaction have explored various factors within the compliance–reliance paradigm to explain operator behavior \cite{dixon2006automation}. Trust \cite{lee2004trust} and risk \cite{mayer1995integrative} play important roles in human dependence on automation. Trust is defined as “the trustor’s willingness to assume risk by delegating responsibility to the trustee” \cite{mayer1995integrative}. Risk is defined as “the level of uncertainty as to whether the decision might lead to significant and/or disappointing outcomes” \cite{sitkin1992reconceptualizing}. Chancey, E. T., et al. proposed a conceptual model that illustrates how system errors impact dependence, with risk moderating the effect of trust \cite{chancey2017}. They found that the false alarm rate affected compliance but not reliance, while the miss rate affected reliance but not compliance. Trust mediated the relationship between false alarm rate and compliance, affecting reliance only within high-risk groups.

\begin{figure}[h]
  \centering
  \includegraphics[width=0.6\textwidth]{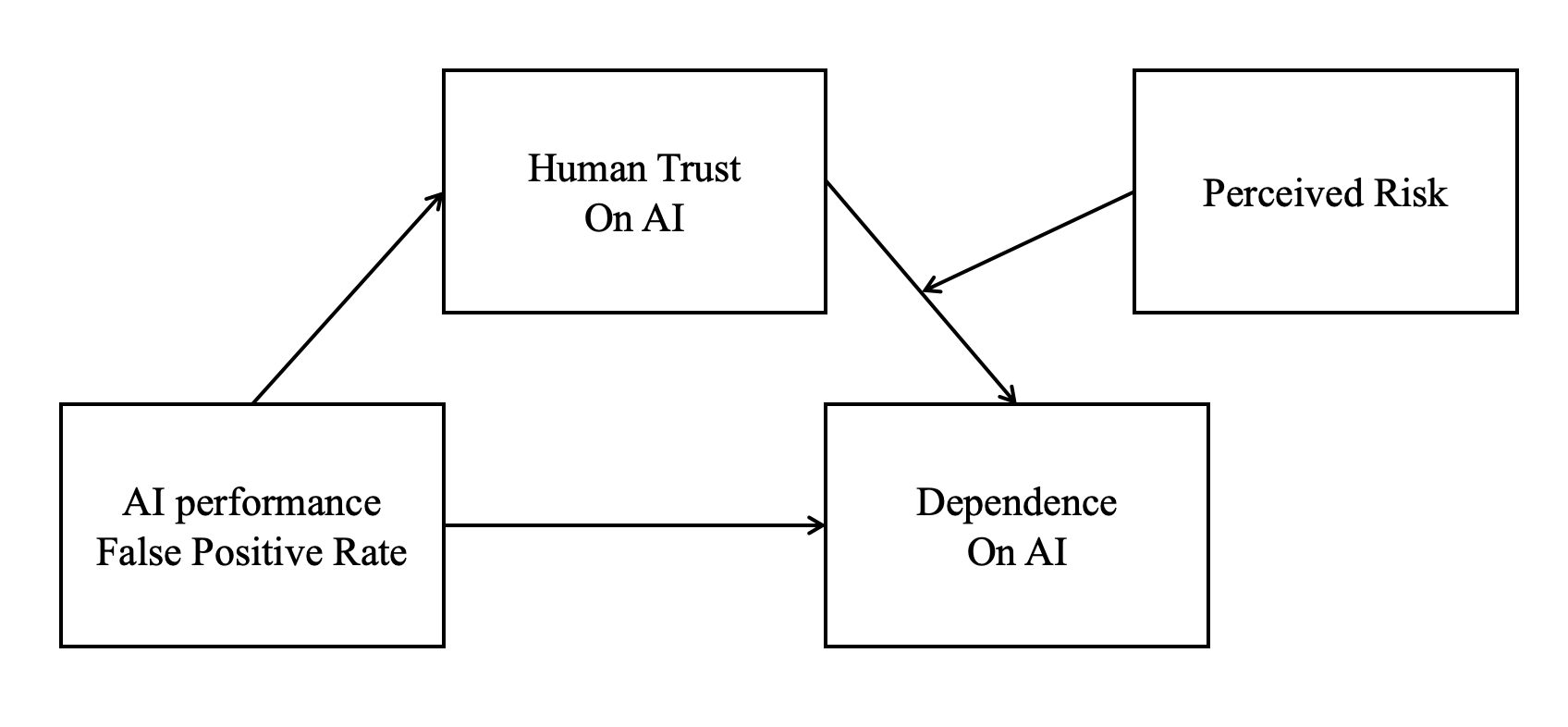}
  \caption{Hypothesized moderated mediation model depicting the relationship between FPR, trust, risk, and dependence.}
  \label{fig:h-model}
\end{figure}

To study the decision making process supported by deepfake detection AI, we chose to adopt Chancey, E. T., et al.'s theoretical framework \cite{chancey2017} to human-AI interaction and proposed our research hypotheses (as shown in Figure~\ref{fig:h-model}).

We interpret false alarms as false-positive errors in the AI’s predictions, where the AI predicts an image to be synthetic when it is real. A missed alarm is seen as a false-negative error, where the AI predicts an image to be real, though it is synthetic.

While AI performance can be assessed through various metrics, our study focuses on the false positive rate (FPR). This choice is based on the fact that in practical applications, users are more likely to notice and respond to system warnings, perceiving possible errors when the system identifies their handwritten text or photos as synthetic. Consequently, we propose the following hypotheses:

\textbf{H1:} If the AI system reveals that it has a lower false positive rate (FPR), this will lead to greater trust in AI.

\textbf{H2:} There will be an interaction between the results predicted by AI and the AI's performance on whether humans choose to align with the AI's prediction.

\textbf{H3:} Trust will moderate the relationship between AI performance and compliance.

\textbf{H4:} Trust will moderate the relationship between AI performance and reliance.

\textbf{H5:} Risk will moderate the mediating effect of trust.

\section{Methods}
To answer our research questions, we conducted an online experiment. The study was approved by the university's IRB. Informed consent was obtained from each participant.
\subsection{Experiment Design}
Based on the hypotheses, we conducted a subjective human experiment to explore how the impact of instructions about AI system performance, the perceived risk of deepfakes, and human trust in the AI system impact human decision making behavior in the context of deepfake detection. We implemented a 2x2 design, manipulating AI performance (high FPR vs. low FPR) and risk level (high vs. low) split-plot design. The risk level was manipulated between subjects, and the AI performance was manipulated within subjects. 
\subsubsection{Experimental task}
\begin{figure}[!t]
    \centering
    \includegraphics[width=1.0\textwidth]{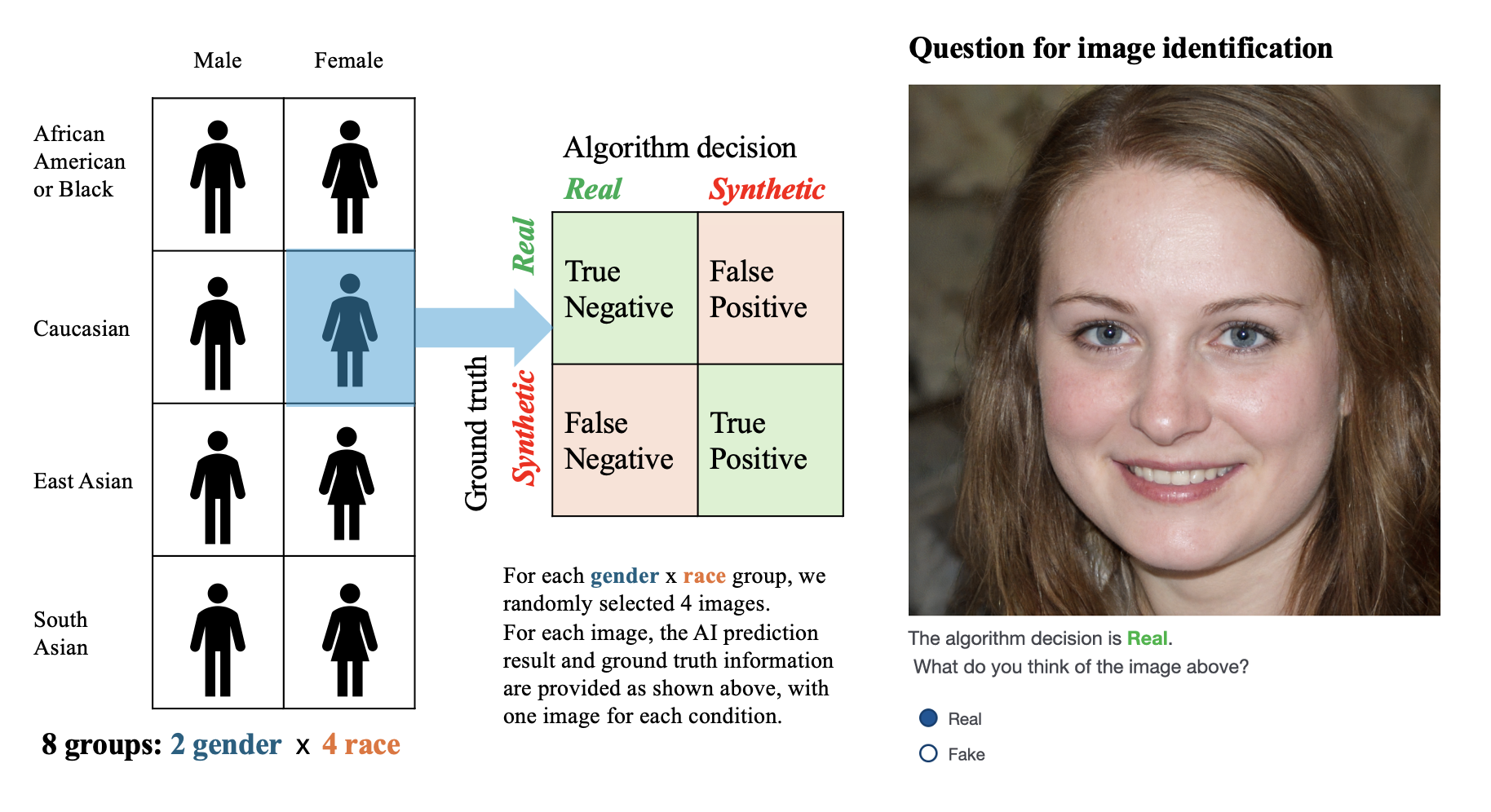}
    \caption{Experimental task description for each AI system. The participants were tasked with determining the authenticity of 32 images.}
    \label{ExpTask}
\end{figure}

As shown in Figure~\ref{ExpTask}, for each AI system, participants were tasked with determining the authenticity of 32 images, with presenting AI prediction results. These images were sourced from a publicly available scientific dataset \cite{nightingale2022ai} and were labeled according to gender and ethnicity. The gender categories included male and female, while the ethnicities represented were Black, White, Middle Eastern, and Asian. From the image sets of eight gender and ethnicity combinations depicting adults, we randomly selected two real and two synthetic images. We ensured that only facial features were visible, excluding jewelry or other body parts.
\subsubsection{Risk Manipulation}
Perceived risk towards deepfakes refers to the degree of uncertainty that confusion between deepfakes and real information could lead to significant and/or disappointing outcomes. To manipulate participants' perceived risk, participants assigned to the high-risk group were provided with two case examples illustrating the harms caused by AI-generated synthetic images, as reported by government agencies \cite{deepfakeReport}. Previous literature has demonstrated that users tend to distrust AI systems due to the political leanings of the companies that develop and operate them \cite{jahanbakhsh2023exploring}. To remove confounders in the experiment, we decided to only provide daily life cases from government reports, such as deepfake kidnapping and cyberbullying, that do not require any other specialized background knowledge and are unrelated to political topics. Participants in the low-risk group proceeded directly to the relevant task without reading any information. Considering that participants’ perceived risk may vary after being exposed to images in the experiment, we evaluate the perceived risk at the end of the experiment.
\subsubsection{AI Performance Manipulation}
This experiment aims to assess whether directly revealing the false positive rate (FPR) affects users' trust and behavior. Before participants began the image identification task in each round, we explained the AI system's performance in making predictions for that round, using examples. A false positive occurs when the AI mistakenly labels a real image as synthetic, and a high FPR suggests that the system may frequently misclassify real images or videos as AI-generated. Although there was no actual AI system in this study, participants were told that there was one. The high-FPR AI was described as having a 30\% false positive rate, while the low-FPR AI had a rate of 3\%. 

It is important to emphasize that although we manipulated the AI's performance, we did not align the probability of errors in predicting the results for 32 images with their false positive rate (FPR) during the experiment. This is because two systems with differing performances can produce similar prediction outcomes on the same set of images, making it difficult for users to fully evaluate the AI's performance based on limited data. Consequently, within each race and gender group, the AI provided one prediction as real and one as synthetic for the two synthetic images. Likewise, for the two real images, the AI also issued one prediction as real and one as synthetic. In summary, for the four images within each race and ethnicity group, the AI's predictions comprised one true positive, one true negative, one false negative, and one false positive.

\subsubsection{Procedure}
\begin{figure}[h]
  \centering
  \includegraphics[width=0.9\textwidth]{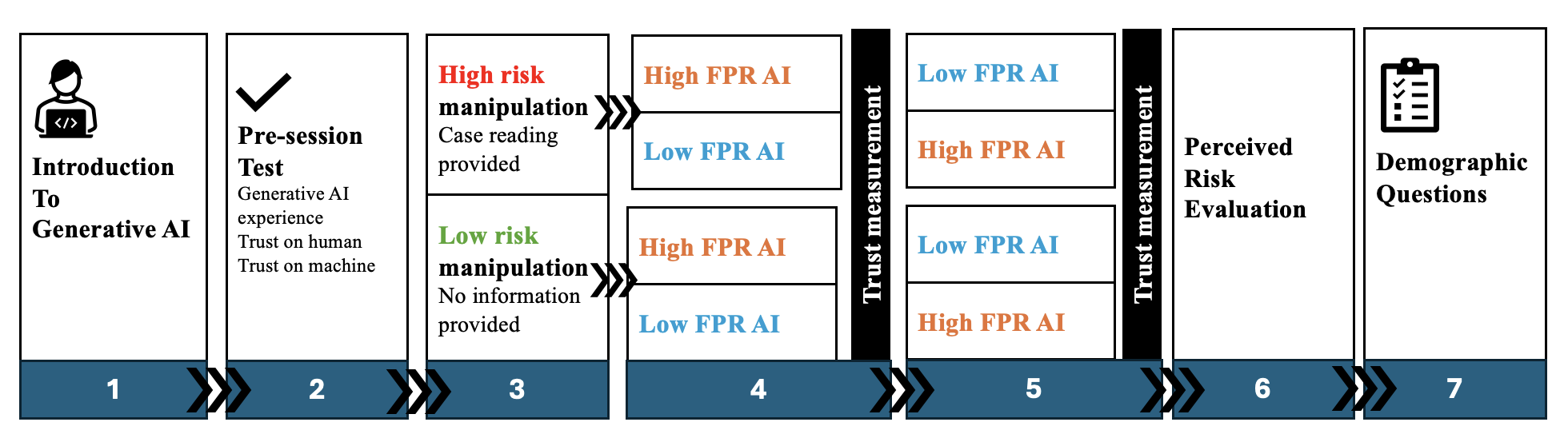}
  \caption{The flow chart depicting the experiment procedure}
  \label{fig:t-f}
\end{figure}

As shown in the figure \ref{fig:t-f}, participants entered the experiment only after reading and consenting to the informed consent form. They first read an introduction regarding generative AI and answered questions related to their experiences with generative AI and synthetic images. Next, they responded to questions assessing their general trust in social and AI systems. Following this, they received the risk manipulation articles. Then, they were provided with instructions on the performance of the first AI system and completed the first part of the experimental task with the assistance of the first AI system. After identifying the authenticity of the 32 images, we adopted existing questionnaires regarding trust in AI to measure participants' trust in the first AI system \cite{chen2023ai}. Then, they were provided with instructions on the performance of the second AI system and completed the task with the assistance of the second AI system and again assessed their trust. To eliminate the bias in trust calibration caused by the order of presentation of the AI systems, the order of presentation for the two AI systems was randomized in each group. After completing both tasks, we used a modified version of the risk questionnaire to measure participants’ perceived risk \cite{chancey2016effects}. Finally, participants answered a series of demographic questions.

\subsection{Participants}
We used G*Power software \footnote{https://www.psychologie.hhu.de/arbeitsgruppen/allgemeine-psychologie-und-arbeitspsychologie/gpower.html} to calculate our minimum sample size based on an f-test with a power of .95 and an error rate below .05, arriving at a requirement of 212 participants. In total, we recruited 400 participants on Prolific \footnote{https://www.prolific.com/} in July and August 2024. The participants were paid based on an hourly rate of $\$8$ which is the minimum wage \footnote{Prolific's Payment Principles: https://researcher-help.prolific.com/en/article/2273bd} for Prolific platform. Considering whether participants passed the attention check question and completed all questions, we finally obtained 390 valid responses. This sample size exceeds our initial calculation, ensuring the sufficiency of our subsequent data analysis.

We restricted the recruitment region to the United States. Among the 390 participants who completed the study, 199 identified as White, 79 as Black, 59 as Asian, 35 as Hispanic/Latino, and 4 as American Indian/Alaska Native, while the remainder chose not to disclose their ethnicity. Additionally, 228 participants identified as female, 147 as male, and 10 as non-binary; the rest opted not to disclose their gender information. Regarding age distribution, 156 participants were between 18–29 years old, 129 were 30–39, 53 were 40–49, 38 were 50–59, and the remaining 12 were over 60. In terms of educational attainment, 335 participants had completed at least some college credits or held a bachelor's degree or higher. Initially, 400 participants were evenly assigned into four experimental groups categorized by false positive rate (FPR) manipulations and risk treatment conditions; of these, 390 completed the study. 

To understand participants’ prior familiarity with deepfake, we collected post-study survey responses on two aspects: where participants had previously encountered deepfake images and their experience with photo editing software.

As shown in Table~\ref{tab:deepfake_exposure} in Appendix, the vast majority of participants across all four groups reported seeing deepfakes on social media platforms (ranging from 77.1\% in Group 4 to 82.5\% in Group 1). A smaller proportion of participants identified news websites or other sources as places they had encountered such content. Notably, approximately 7–8\% in each group chose not to answer this question.

In terms of participants’ technical experience, Table~\ref{tab:photo_editing_experience} in Appendix reveals a broad spectrum of familiarity with photo editing tools. Between 25.0\% and 45.4\% of participants in each group reported never having used such tools, while another 36.1–44.8\% had some prior experience. Group 3 stood out as having the highest proportion of participants with substantial photo editing experience (18.0\%), whereas Group 1 had the highest proportion of completely inexperienced participants (45.4\%). Only a small number in each group indicated extensive expertise or declined to answer.

\subsection{Analysis}
We first calculated the validity of the trust \cite{chen2023ai} and risk questionnaires \cite{chancey2016effects}. The Cronbach's alpha for the questions measuring trust was 0.89, and for the questions measuring perceived risk, it was 0.84. This indicates that these questions demonstrate good reliability and are suitable for research. All items regarding trust and risk were measured on a 7-point Likert scale, ranging from 1 = strongly disagree to 7 = strongly agree. We averaged the responses to these questions to represent an individual's trust in AI systems and perceived risk scores.

We define "compliance" as the proportion of synthetic images identified as synthetic, out of the total number of predicted synthetic images. We define "reliance" as the proportion of real images identified as real, out of the total number of predicted real images. These values range from 0 (minimum) to 1 (maximum).

We used R to test the hypothesis. We applied mixed ANOVA from the afex package \cite{singmann2015package} to examine the main effects and interactions among dependent measures. In the mediated moderation analysis, we applied Model 14 from the process package \cite{hayes2017introduction} to test whether risk modifies the degree to which trust mediates the relationships tested. We set 5,000 bootstrap resamples.

\section{Results}
\subsection{Manipulation check}
We manipulated participants' perceived risk at the beginning of the experiment. At the end of the experiment, we measured users' perceived risk (See Appendix~\ref{fig:risk_perception_survey}) regarding the malicious outcomes caused by synthetic images. Through individual T-test, we found no significant difference between the group which received high risk manipulation (M = 4.78, SD = 1.36) and the low-risk groups (M = 4.90, SD =1.40). Considering that most participants had already encountered and were aware of synthetic images before our experiment, merely presenting the negative consequences of synthetic images does not significantly heighten people's awareness of such risks.

\subsection{Hypothesis testing}
\subsubsection{Effects on trust}

We applied paired T-test to compare the two trust measurements on the participants towards the high-FPR AI system and the low-FPR AI system. Supporting H1 ($T = 2.25$, $p < .05$), there was a significant difference between human trust in the high-FPR AI system (M = 3.62, SD = 1.36) and the low-FPR system (M = 3.75, SD = 1.41). An effect size of $d = 0.09$, as measured by Cohen’s d, reflects a negligible magnitude of difference. No other significant effects were observed. This indicates that presenting information about AI with a lower false positive rate increased participants’ trust in AI. Our manipulation of FPR was effective. However, considering that our Likert questions used for trust measurement scale ranged from 1 = strongly disagree to 7 = strongly agree. The results suggest that our participants did not trust either of our two systems.

Before the experimental tasks, we assessed participants' trust in other humans (M = 5.34, SD = 0.90) and their trust in general AI (M = 3.61, SD = 1.20). The results of paired T-test ($T = 22.38$, $p < .001$) and Cohen’s d effect size of $d=1.6$ indicated that participants showed significantly lower levels of trust in AI compared to humans, which was also reflected in their distrust in the two deepfake detection AI systems during the experiment.

\subsubsection{Effects on human decision}
The participants were randomly assigned to receive one of the risk manipulation treatments. Each participant experienced two different AI systems, and in each AI system, we provided two types of prediction results: either a real image or a synthetic image. Therefore, we chose to conduct a mixed ANOVA to test H2. 

The FPR, performance of AI($F(1, 388) = 11.16$, $p = .0009$, $\eta_{\text{p}}^{2} = 0.03$), and predictions given by AI ($F(1, 388) = 59.91$, $p < .001$, $\eta^2_{\text{p}} = 0.13$) have significant main effects on whether humans will follow the decisions of the AI system, which suggested that each factor independently shaped how humans identify deepfake images with AI assistant.

In addition to the main effects, we observed several significant interactions. There was a significant interaction between perceived risk and FPR ($F(1, 388) = 7.97$, $p = .005$, $\eta^2_{\text{p}} = 0.02$), indicating that whether participants chose to follow the decisions of AI systems with different FPR also depended on whether they perceived the risk to be high or low. We also observed a significant interaction between FPR and the predictions provided by the AI ($F(1, 388) = 8.59$, $p = .004$, $\eta^2_{\text{p}} = 0.02$), indicating that the impact of FPR on human decision making differed depending on the predictions given by the AI. The interaction between perceived risk and the prediction results provided by AI alone was not significant ($F(1, 388) = 0.85$, $p = .358$). Notably, there was a a three-way interaction between perceived risk, FPR, and prediction given by AI system($F(1, 388) = 21.36$, $p < .001$, $\eta^2_{\text{p}} = 0.05$), highlighting that perceived risk and different decisions given by AI systems with different FPRs jointly influenced whether participants followed the AI's decision.

\begin{figure}[h]
  \centering
  \includegraphics[width=0.9\textwidth]{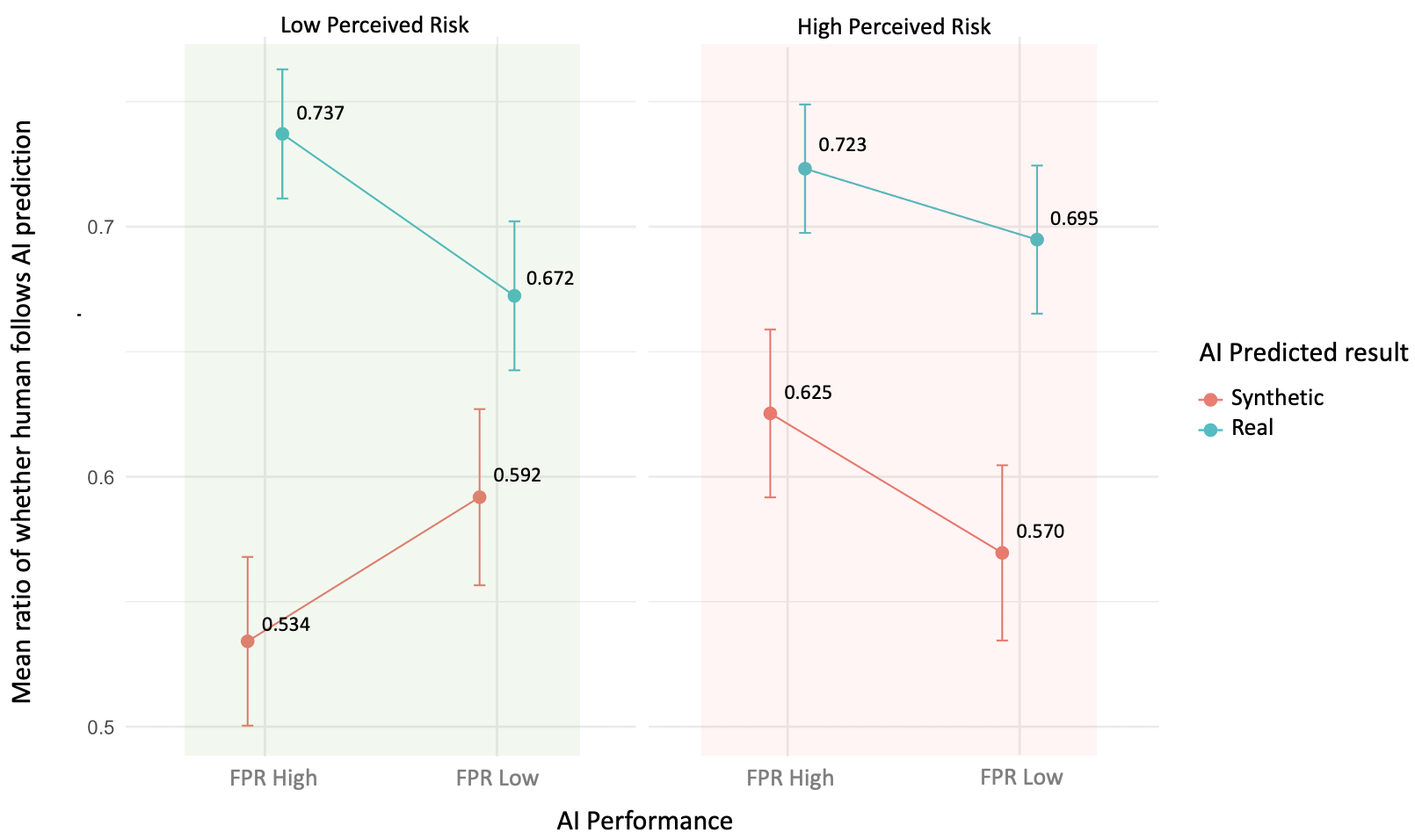}
  \caption{Mixed ANOVA result with 95\% Confidence Interval}
  \label{fig:anova}
\end{figure}

To better understand the significant three-way interaction, we explored the FPR × prediction results interaction at each level of perceived risk. As shown in the figure \ref{fig:anova}, in the group with manipulation of low perceived risk, when the image is predicted as synthetic, the ratio of humans following the decisions of the low-FPR AI system (M = 0.59, SE = 0.02) was significantly higher than the high-FPR AI system (M = 0.53, SE = 0.02,  $p = .001$). It is not surprising, reflecting skepticism toward high-FPR AI making synthetic judgments. Conversely, when the image is predicted as real, the ratio of participants following the decisions of the high-FPR AI system (M = 0.74, SE = 0.01) was significantly higher than the ratio of human following the decisions of the low-FPR AI system (M = 0.67, SE = 0.01, $p = .0001$).

In the group with manipulation of high perceived risk, the pattern was different: When the AI predicted an image as real, there was no statistically significant difference in whether humans followed the predictions provided by the high FPR AI system (M = 0.72, SE = 0.01) or the low FPR AI system (M = 0.70, SE = 0.01, $p = .22$). When the image is predicted as synthetic, participants became more likely to follow the high FPR AI system (M = 0.63, SE = 0.02) and less likely to follow the low FPR AI system(M = 0.57, SE = 0.02, $p = .0015$). It indicates that high perceived risk may induce participants preferring over-detection of synthetic images (false positives) to under-detection of potential threats (false negatives).

\subsubsection{Mediation Analyses}
To explore the underlying mechanisms by which participants follow AI predictions, we applied Model 14 in PROCESS for R \cite{hayes2017introduction}.

In the case of compliance, when the AI system identified the image as synthetic and the human decision was consistent with the AI’s prediction, the hypothesized model was not statistically significant ($F(4, 775) = 1.68$, $p = 0.15$). The FPR levels did not have a statistically significant direct effect on humans’ trust in AI ($B = -0.128$, $p = .196$). Similarly, trust in AI did not significantly predict compliance ($B = 0.059$, $p = .092$). No significant direct effect of perceived risk on compliance was observed ($B = 0.045$, $p = .051$). Furthermore, the interaction between trust and perceived risk, the moderated mediation pathway, was not statistically significant ($B = -0.009$, $p = .155$). Thus, we rejected H3.

\begin{figure}[h]
  \centering
  \includegraphics[width=0.6\textwidth]{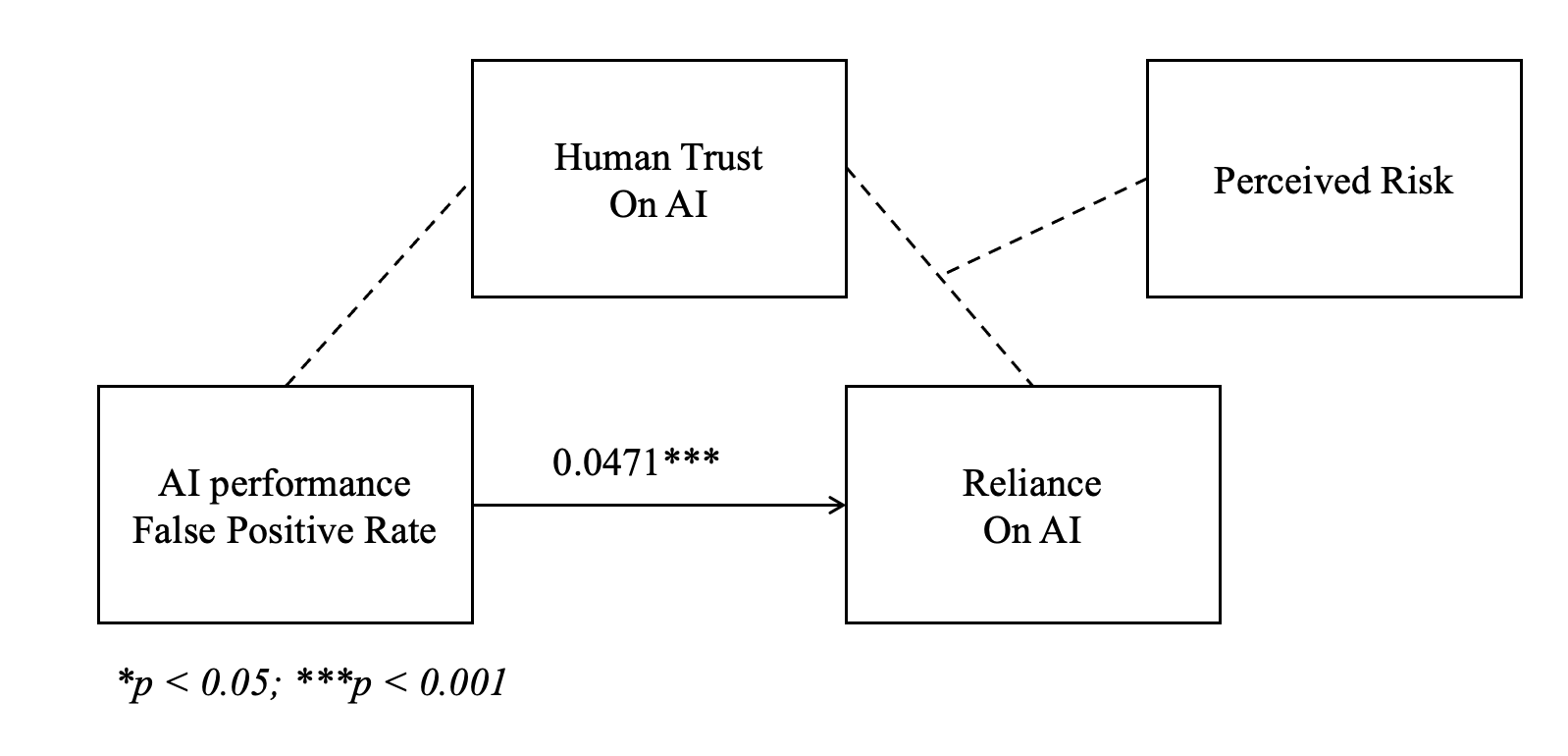}
  \caption{Hypothesized moderated mediation model depicting the relationship between FPR, trust, risk, and dependence.}
  \label{fig:MM}
\end{figure}

In the case of reliance, when the AI system identified the image as real and the human decision was consistent with the AI’s prediction, as shown in the figure \ref{fig:MM},  we only observed the direct effects of AI performance (FPR) on reliance ($B = .0471$, SE = .01, 95\%, CI: [0.02, 0.07], $p < .001$). FPR levels did not significantly predict trust in AI ($B = -0.13$, $p = .20$). Trust did not significantly predict reliance ($B = -0.0015$, $p = .96$). There was no significant interaction between trust and risk ($B = 0.0012$, $p = .81$).  Thus, we rejected H4 \& H5.

\section{Discussion}
Through online experiments, we investigated how the performance of AI systems, trust, and risk impact users' reliance and compliance when completing tasks beyond their capabilities. In this section, we will first discuss the challenge in current system designs when AI assists people in decision making based on our findings. Next, we will explore design implications for enhancing human-AI decision making to combat the threats of deepfake.

\subsection{Trust: Impact of Performance Information on Trust in AI Systems}
Our findings validate previous studies showing that people's trust in AI systems can be influenced by the AI system’s performance. In the experiment, we directly informed participants about the AI's performance and provided corresponding explanations. When participants were told that the AI system had higher performance, their trust in the system increased. However, in our online experiment, although participants increased their trust in the low-false-positive-rate AI system, their average trust level remained in the distrust range, similar to their general machine trust levels measured prior to the experiment. Consequently, our findings reject the hypothesis and differ from previous studies \cite{chancey2017}. When users generally distrust a system, variations in trust within this range do not affect their dependence on AI.

\subsection{Risk: Role of Perceived Risk in Trust and Compliance-Reliance} 
Based on our manipulation check on perceived risk at the end of the study, our participants do not appear to be concerned about the risks posed by deepfakes. Even among those who are concerned, they remain a minority overall. It is aligned with some previous field studies. Based on our results, it remains uncertain whether risk moderates the mediating effect of trust on compliance or reliance. Trust is often defined as the willingness to rely on a machine or an AI, even if it may expose humans to some vulnerabilities. However, previous literature in the area of interpersonal interaction \cite{sitkin1992reconceptualizing, lyons2012human} suggested that risk is a critical factor in whether individuals accept others' decisions and bear responsibility for potential failures and vulnerabilities. Given that most of our participants expressed distrust of any general AI system, as well as the two deepfake detection AIs in our experiment, they were unwilling to take on the risk of AI prediction failure themselves. Thus, in our experiment, we cannot observe how perceived risk moderates trust.

\subsection{Compliance-Reliance} 

We observed a significant three-way interaction between perceived risk, AI performance (measured via false positive rate, FPR), and the type of prediction provided by the AI. Specifically, when the AI predicted an image as synthetic, the ratio of whether humans followed the AI’s prediction, referred to as compliance, varied by perceived risk level. In the group with manipulation of high perceived risk, our data indicated that participants showed more compliance with high FPR AI systems. It may indicate a preference for systems that prioritize caution, even at the cost of over-detection. Conversely, in the group with manipulation of low perceived risk, participants showed more compliance with low FPR AI systems, suggesting a stronger sensitivity to AI performance. These contrasting patterns highlight how perceived risk influences users’ trust and decision strategies in human-AI interaction, with participants calibrating their compliance on AI based on both AI system factors and task factors. The findings are consistent with previous literature, showing that false alarm rates affect user compliance with automation\cite{chancey2017,alahmadi202299,dixon2006automation}.

When the AI predicted an image as real, the ratio of whether humans followed the AI’s prediction, referred to as reliance, also varied based on perceived risk. In the group with manipulation of high perceived risk, there was no statistically significant difference in reliance between the high and low FPR AI systems. In contrast, participants with manipulation of low perceived risk demonstrated more reliance on the high FPR AI. Our results differ from previous research \cite{chancey2017} that the false alarm rate is not related to reliance. As in our study, only the false positive rate was explicitly communicated to participants, without further contextualization, participants may have inferred that a high FPR AI system would be less likely to misclassify synthetic content as real.

\subsection{Challenges and Design Implications}
Based on our experiment findings, we identify three challenges in mitigating the harms of deepfakes: (1) distrust in deepfake detection AI models, (2) misperceptions on the performance of AI systems, and (3) insufficient risk awareness regarding the potential impacts of deepfakes. We will elaborate on these three points in this section and discuss possible design implications.

\subsubsection{Distrust in deepfake detection AI models} 
Our experiment's context is identifying synthetic content, which is often closely linked to misinformation. Previous studies have tested people's trust in systems that moderate misinformation and found that users tend to question these systems based on their political stance, conspiracy beliefs, and other such factors \cite{jahanbakhsh2023exploring}. Although we reduced the association between AI-generated content and false political news in our risk manipulation, participants in our study still showed distrust toward the manipulated system, echoing findings in misinformation detection literature. It might imply that the distrust on those system is not a rare case. As shown in our experiment where participants distrust the deepfake detection system, they are less likely to follow the AI prediction as synthetic than the AI prediction as real. This pattern occurs regardless of the system's actual performance, meaning that even when these detection systems perform well, users are inclined to ignore deepfake warnings if they don’t trust the system, even if they lack the ability to assess the content’s authenticity on their own. Such situations may amplify the threat posed by synthetic data.

\textbf{Design Implication: Design for Trust}: Design should serve as a bridge between existing AI technologies and the users who interact with them. Future research should investigate users’ mental models and concerns surrounding AI-assisted deepfake detection, especially as shaped by broader sociopolitical contexts. Understanding these dimensions can inform more trust-sensitive design strategies. Moreover, in some cases, AI performance metrics may not be persuasive to users. Future deepfake detection AI system design should explore to incorporate additional forms of trust-building except for the AI model itself, such as third-party audits and transparency around training data.

\subsubsection{Misperceptions on the performance of AI systems}
Our findings suggested that participants’ dependence, including reliance and compliance, on the AI system was impacted by both the AI’s prediction and the contextual information provided by researchers, such as descriptions of deepfake risks and the system’s reported performance. Participants were not blindly following the AI. They engaged in situated trust calibration, dynamically weighing the known limitations of the AI (FPR), the AI's current decision (synthetic vs. authentic), and their perceived risk of deepfakes. 

Across all perceived risk groups, we observed that participants tend to perceive images identified by the system as "real" as more real. Their reliance on both AI systems was significantly higher than compliance. This aligns with what prior literature refers to as the "implied truth effect" \cite{pennycook_implied_2020}. It refers to the tendency for warning labels on clearly false content to make other types of (potentially subtler) misinformation seem more credible. Users may misperceive the absence of a warning as a signal of truth.

Although our experiment included both “real” and “synthetic” labels, we only communicated information about the FPR. This partial and directional communication between the designer and the user likely contributed to participants’ misperception of the AI system's performance. Communicating AI transparency and performance effectively, especially regarding its limitations,emerges as a critical challenge in mitigating the harm of deepfakes.

Additionally, given that prior literature has measured human ability and confidence in identifying deepfakes, our experimental design only focused on how participants calibrated their trust across two AI systems with differing performance levels. As previous studies have shown, human dependence in AI is not only grounded in statistically measured performance \cite{salimzadeh2024dealing}, but also in whether the system’s output feels aligned with one’s own intuition or judgment \cite{ma2024you,ma2023should}. This suggests that, even if humans lack the ability to accurately identify deepfakes, they may still apply a form of subjective evaluation to the AI’s prediction. It might explain why some participants rely more on the higher-risk AI system. However, those subjective judgments were not fully captured in our current study, a limitation worth noting.

In the context of deepfake detection, such misalignment between user perception and AI performance may potentially trigger the "backfire effect", where individuals, when confronted with information that contradicts their existing beliefs, become even more entrenched in those beliefs rather than revising them \cite{wood2019elusive}. In our context, this could manifest as growing resistance or skepticism toward deepfake detection AI systems altogether.

\textbf{Design Implication: Design for two-way communication between users and AI}: Designers should take care in how AI communicates decisions with humans, the ways it presents results, and what types of results it shows. Such design decisions should be informed by specific user groups and by the collaborative task between AI and humans. Context-aware explanations and adaptive interfaces are necessary to highlight various aspects of AI system performance in relation to the user’s situational context and decision making patterns. For example, AI system design may allow users to express their level of agreement or uncertainty about predictions, which could surface humans' implicit beliefs and prompt AI to provide more plausible explanations for predicted results and its performance.

\subsubsection{Insufficient risk awareness regarding the potential impacts of deepfakes} 

According to our final manipulation check questionnaire on risk perception, most participants did not think deepfakes are a serious threat. We found that participants showed a converse compliance pattern under different levels of risk when the AI predicted an image to be synthetic; participants were more likely to follow the low-FPR AI system than the high-FPR AI system in the group with manipulation of low perceived risk. However, this pattern reversed in the group with manipulation of high perceived risk. It suggests that perceived risk alters the threshold of acceptable error made by AI, with participants preferring errors on the side of false alarms (i.e., false positives) rather than risk missing an alarm (false negatives). In other words, under perceived risk, people appear to align with AI systems that share a more precautionary stance, even if that entails more frequent false alarms. From another perspective, as discussed in the previous section, it also shows that the dependence between humans and AI is dynamic. Considering the potential risks of deepfakes, AI should adjust its communication strategy based on the user's risk preference towards deepfakes.

\textbf{Design Implication: Support perceived risk-oriented Adaptive Communication}: The deepfake detection AI design should not only focus on how to improve user awareness of deepfakes, but also consider how to adjust the AI model itself as users' perceived risk increases. Future AI systems could benefit from adaptive messaging strategies that adjust based on contextual cues or longitudinal user profiles. For example, AI systems may support depicting users' perceived risk towards deepfakes based on their behavior. Based on the user profiles, AI could provide risk-calibrated support to guide users appropriately assess and accept AI predictions.

\section{Limitations and Future Work}
To test participants' compliance/reliance on AI, we recruited participants randomly from the Prolific platform, where users generally have a certain level of digital literacy. Future research should consider the marginalized online populations, such as individuals with low digital literacy or individuals from rural areas.

Most participants were familiar with and had used generative AI, which made our risk manipulation methods less effective with this group. The risk manipulation materials provided in the experiment may lead to different interpretations of the same case on potential harm depending on the reader's perspective as a stakeholder. Future studies should consider tailoring manipulations to participants’ own experiences to effectively influence human awareness of deepfakes.

Additionally, our participants generally presented a distrust of AI or machines, meaning our conclusions may only apply in contexts where people are skeptical of AI. Our conclusions may only demonstrate that in situations of distrust, changes in user trust do not affect their decision making. Future experiments could explore whether trust and distrust, or variety in trust when users show trust in AI, will impact their willingness to adopt AI decisions in tasks that humans show less ability to do.

Our participants were exclusively from the U.S.. Given that laws and regulations regarding deepfakes and generative AI vary across countries \cite{umbach_non-consensual_2024}, caution should be exercised when generalizing these results across cultures. More experimental studies involving diverse cultural groups are needed to understand the extent to which these findings are generalizable.

\section{Conclusion}
Synthetic content generated by AI has become an increasingly significant challenge to cybersecurity, personal privacy, and public trust. In the context of AI-supported synthetic image identification, our study explored how AI performance influences human trust and decision making. Through a human subject experiment (N = 400), we found that participants were more likely to believe images labeled as real by the AI when aware of the AI’s tendency to misclassify real images as synthetic. Although our research aligns with previous studies showing that user trust in AI is sensitive to AI performance, when participants exhibit low general trust in AI, changes in their trust level do not impact their willingness to follow AI decisions. Our study provides insights to further support the development of transparent, explainable AI systems that strengthen public resilience against deepfakes.

\section{Acknowledgments}
This work was supported by the National Science Foundation Award no.2247723. 

\clearpage
\bibliographystyle{ACM-Reference-Format}
\bibliography{sample}

\clearpage
\appendix

\section{Pre-Study Survey Responses Distribution Among 390 Participants}

\begin{figure}[H]
    \centering
    \includegraphics[scale=0.5]{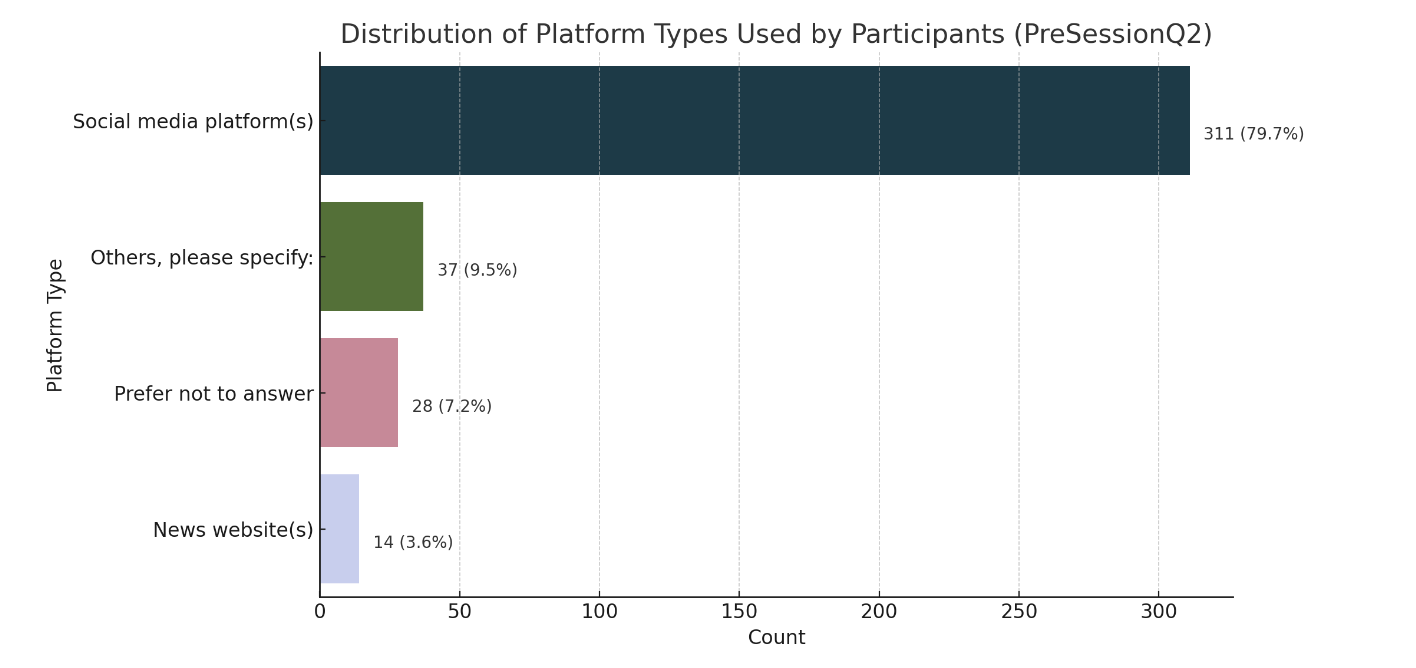}
    \caption{Platform Types Mentioned by Participants in Pre-Study Survey}
    \label{platform}
\end{figure}

\begin{figure}[H]
    \centering
    \includegraphics[scale=0.5]{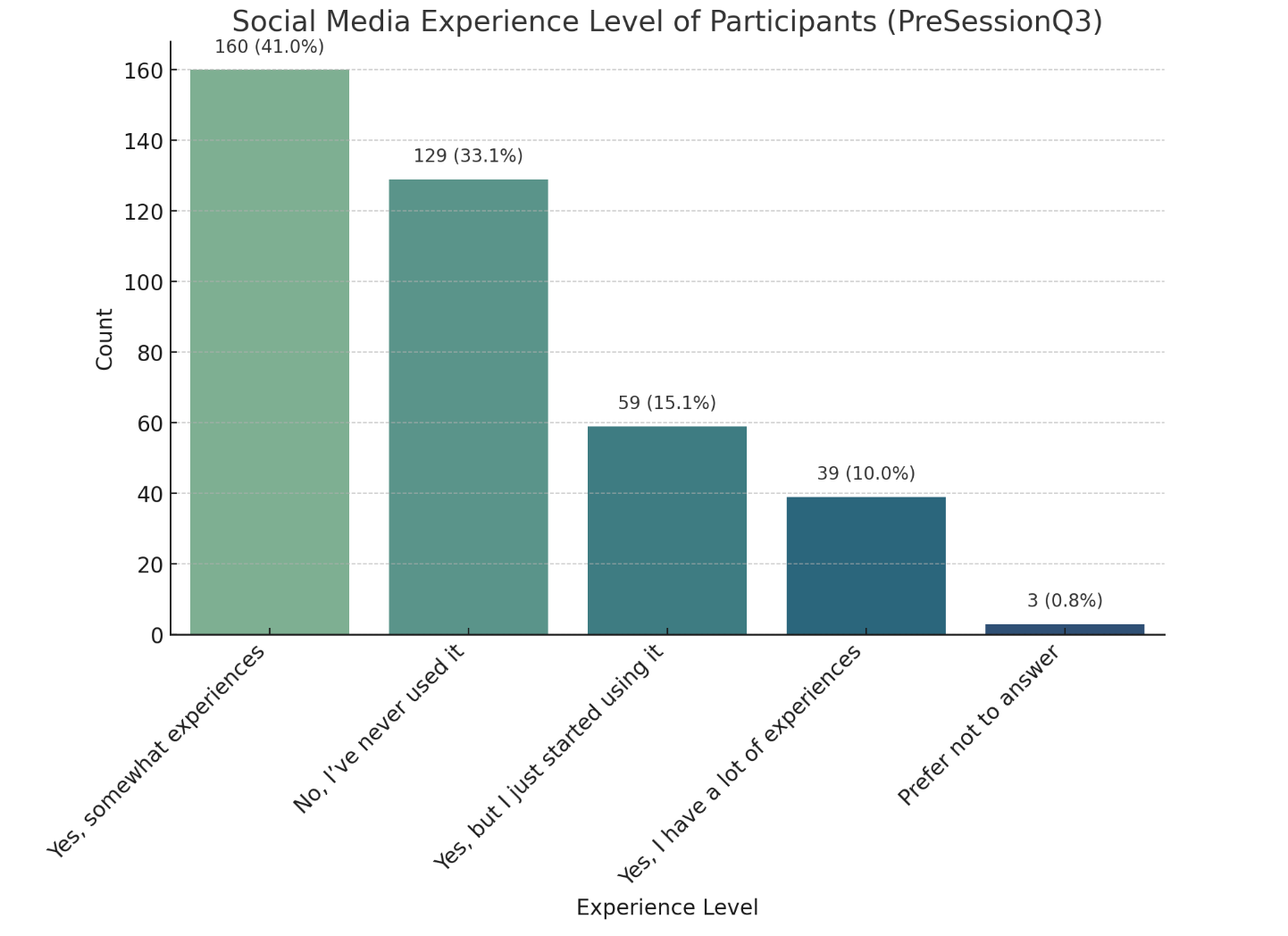}
    \caption{Social Media Experience Level of Participants}
    \label{media_exp}
\end{figure}

\section{Demographic Distribution Among 390 Participants}

\begin{figure}[H]
    \centering
    \includegraphics[scale=0.45]{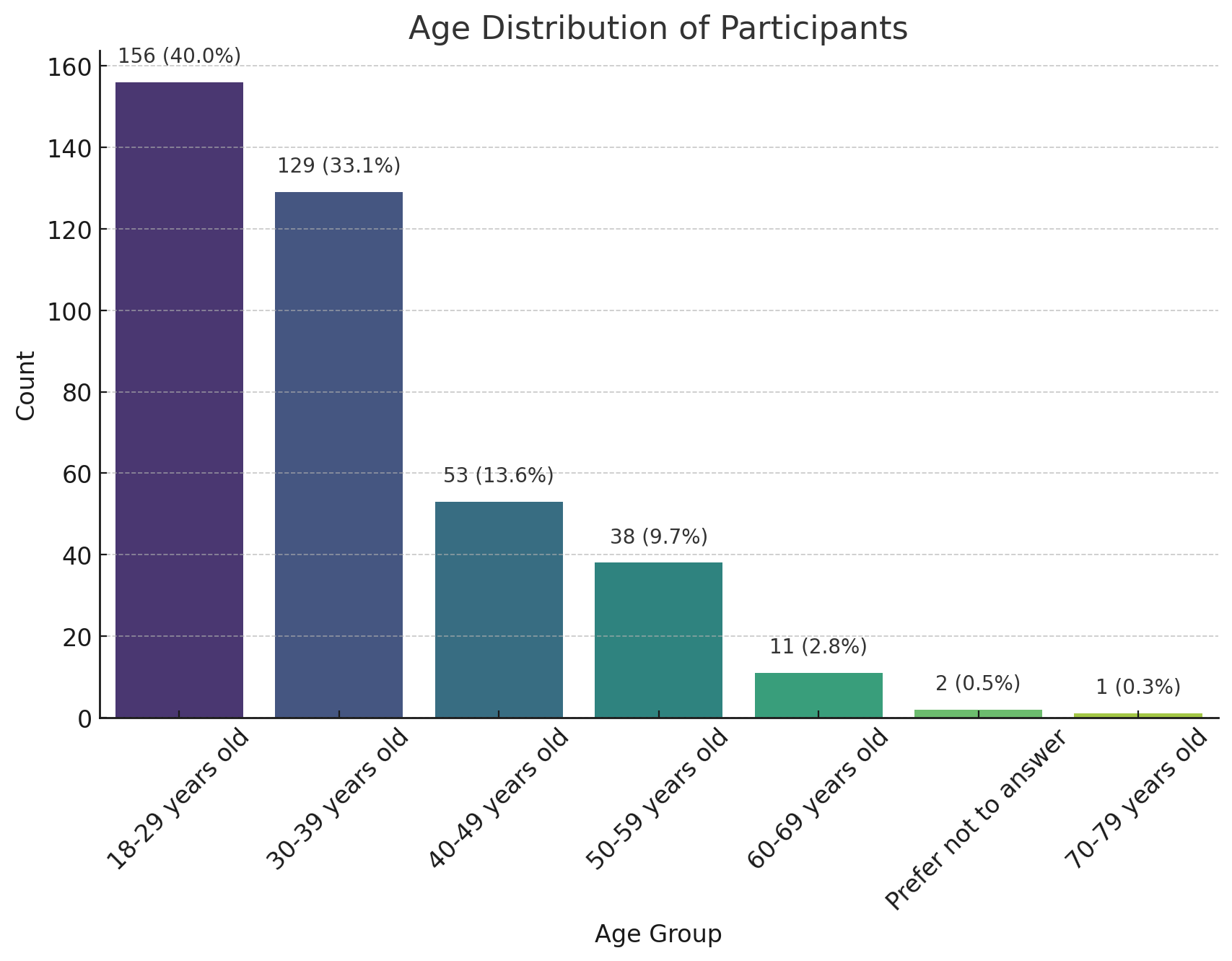}
    \caption{Age Distribution}
    \label{age}
\end{figure}

\begin{figure}[H]
    \centering
    \includegraphics[scale=0.45]{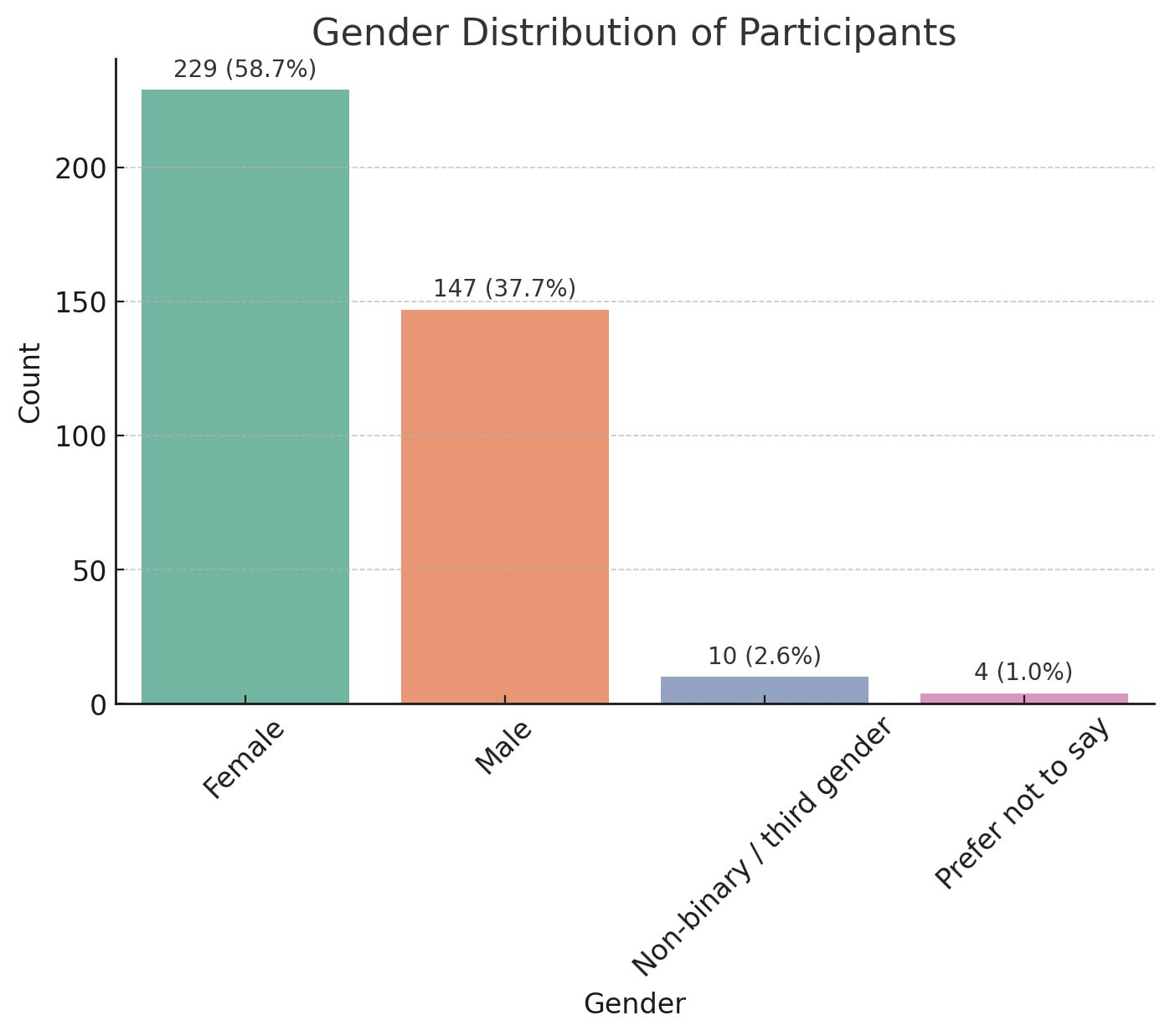}
    \caption{Gender Distribution}
    \label{gender}
\end{figure}

\begin{figure}[H]
    \centering
    \includegraphics[scale=0.45]{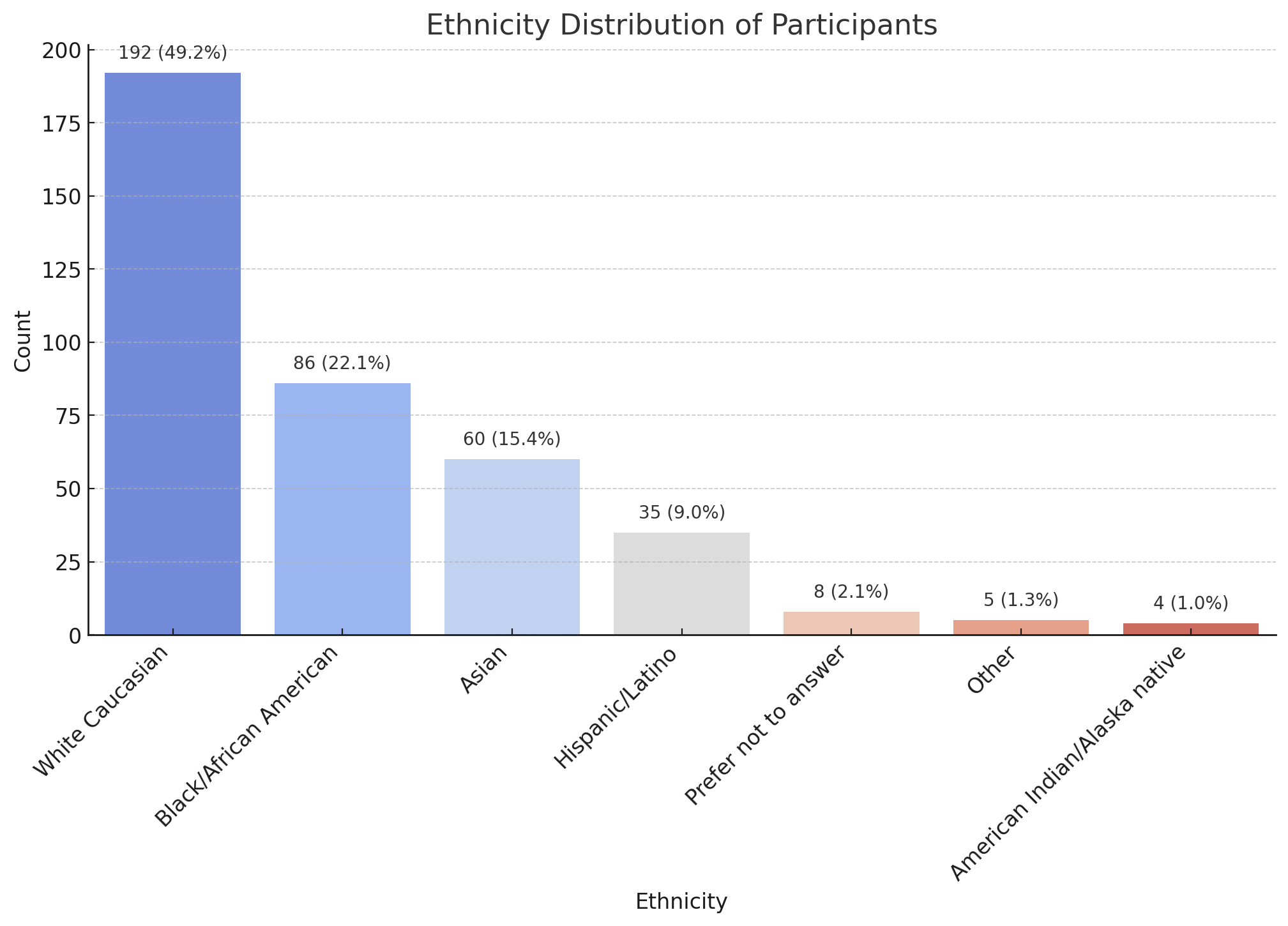}
    \caption{Ethnicity Distribution}
    \label{eth}
\end{figure}

\begin{figure}[H]
    \centering
    \includegraphics[scale=0.45]{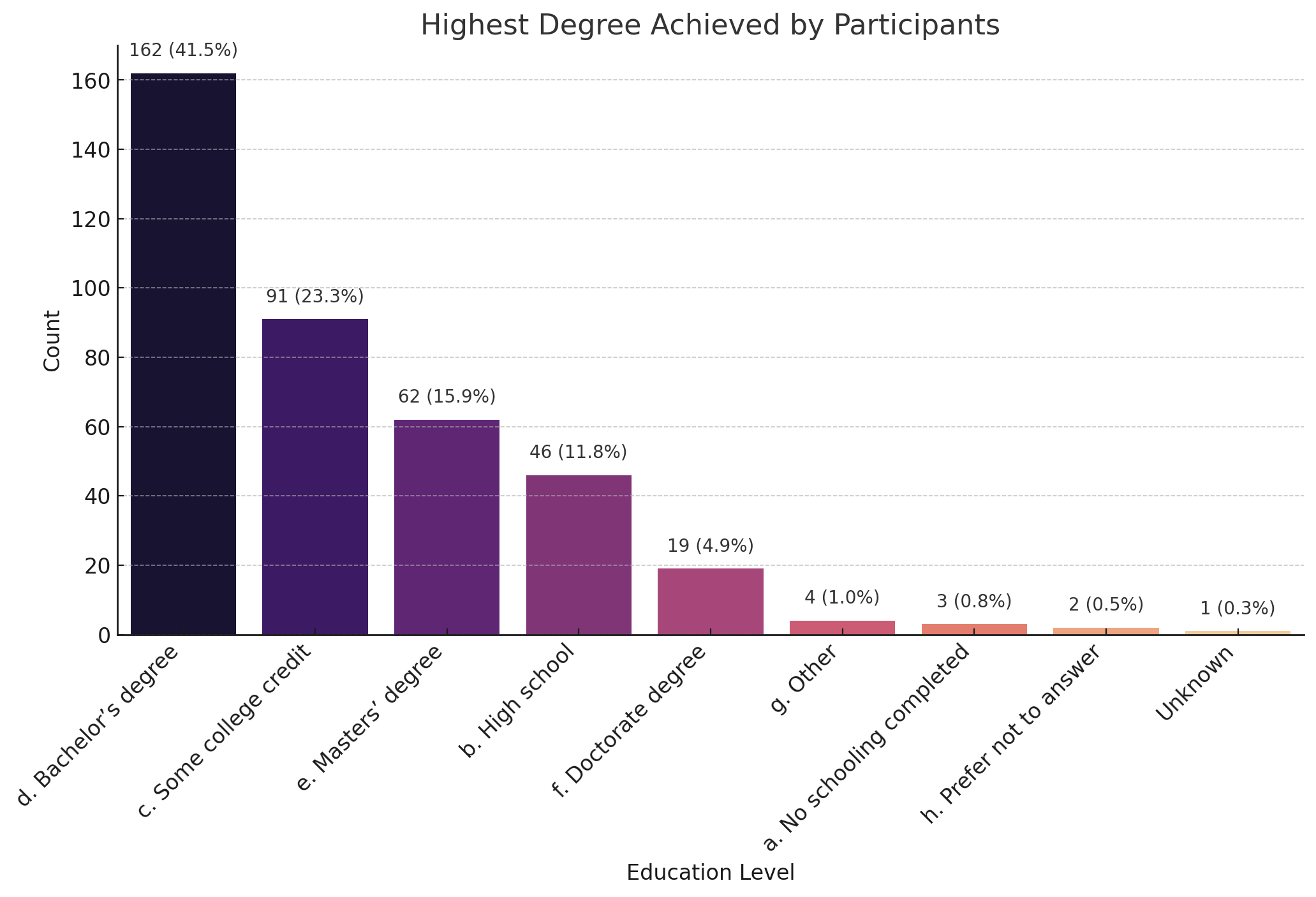}
    \caption{Education Level Distribution}
    \label{edu}
\end{figure}

\section{AI Performance (FPR Level) Manipulation}
\subsection{High FPR 30\%}

To help you with identifying these images are real or synthetic, the AI's prediction decision will be provided.

\textbf{Please note that the synthetic image detection AI will make erroneous decisions.}

A false positive occurs when the AI mistakenly identifies a real image as a synthetic image. If the false positive rate is high, it means that the system may incorrectly identify real videos or images as AI-synthetic.

AI system was mentioned that assists in distinguishing between synthetic and real images, with a false positive rate of \textbf{30\%}. This means that while it makes errors, it rarely misclassifies real images as fake in the mistakes they do make. For example, among \textbf{100} photos identified as fake, ONLY \textbf{30} of them are actually real. 

Please identify each of the following images are real or fake as per your discretion.

\subsection{Low FPR 3\%}

A false positive occurs when the AI mistakenly identifies a real image as a synthetic image. If the false positive rate is high, it means that the system may incorrectly identify real videos or images as AI-synthetic.

AI system was mentioned that assist in distinguishing between synthetic and real images, with a false positive rate of \textbf{3\%}. This means that while it makes errors, it rarely misclassifies real images as fake in the mistakes they do make.

For example, among \textbf{100} photos identified as fake, \textbf{3} of them are actually real.

Please identify each of the following images are real or fake as per your discretion.

\section{Risk Manipulation Presented for High Risk Group}

However, when generative AI is used to create believable, realistic videos, pictures, audio, and text of events that never happened for malicious purposes, those products are called “deepfake” and represent serious risk.

To help participants understand how a potential threat might arise, and what that threat might look like, we considered a number of deepfake scenarios specific to the arenas of society, and national security.

\textbf{Scenario 1. National Security \& Law Enforcement Scenario : Deepfake Kidnapping}

A criminal gang operating in a tourist location in Mexico conducts targeted and opportunistic fraud schemes against victims using synthetic images and video to depict someone in a situation of captivity. The malign actors wouldn’t actually kidnap someone but would use images and information they find either online or from a stolen device to conduct the fraud scheme. The malign actors would then contact the family of the target and demand ransom. They would show the “proof of life” of the victim in a hotel room, possibly bound and blindfolded. The malign actors could also send follow-up images of the victim with indications of injury to place more pressure on the victim’s family to pay the ransom. The victims themselves would not be in direct harm and may be completely unaware of this.

\textbf{Scenario 2. Society Scenario: Cyberbullying}

In March 2021, international news brought to light an incident once charges were filed that involved using alleged deepfake as a cyberbullying tactic. In Pennsylvania, a mother allegedly manipulated images and videos of her daughter’s cheer squad teammates. These alleged deepfakes showed members of the cheer squad drinking, vaping, and posing nude, all actions that could get them cut from the cheerleading squad. Several of the victims came forward about the cyberbullying and one victim claimed the mother went as far as encouraging suicide, furthering the harassment outside of the alleged deepfakes. However, by May 2021, the deepfake accusation had been abandoned as it could not be proven that the video evidence was falsified. Synthetic media researchers noted that the videos did not carry traditional manipulated signatures such as artifacts around the face to suggest it had been altered, but it did have realistic details that would be difficult to fake such as the vapor cloud one individual exhaled. Multiple digital forensics experts stated the videos appeared to be authentic, and it was unlikely they were actually deepfakes.

\begin{table}[H]
    \centering
    \small
    \caption{Perceived Risk Survey}
    \label{fig:risk_perception_survey}
    \begin{tabular}{|p{3cm}|p{1.2cm}|p{1.2cm}|p{1.2cm}|p{1.2cm}|p{1.2cm}|p{1.2cm}|p{1.2cm}|}
        \hline
        \textbf{Statement} & \textbf{Strongly disagree (1)} & \textbf{Disagree (2)} & \textbf{Somewhat disagree (3)} & \textbf{Neutral (4)} & \textbf{Somewhat agree (5)} & \textbf{Agree (6)} & \textbf{Strongly agree (7)} \\
        \hline
        The consequences of poorly identifying real images from deepfake images are substantial. & \(\circ\) & \(\circ\) & \(\circ\) & \(\circ\) & \(\circ\) & \(\circ\) & \(\circ\) \\
        \hline
        The overall risk of poorly identifying real images from deepfake images on these tasks is high. & \(\circ\) & \(\circ\) & \(\circ\) & \(\circ\) & \(\circ\) & \(\circ\) & \(\circ\) \\
        \hline
        Overall, I would label the consequences of performing poorly on identifying real images from deepfake images as something negative. & \(\circ\) & \(\circ\) & \(\circ\) & \(\circ\) & \(\circ\) & \(\circ\) & \(\circ\) \\
        \hline
        I would label the consequences of performing poorly on identifying real images from deepfake images as a significant loss. & \(\circ\) & \(\circ\) & \(\circ\) & \(\circ\) & \(\circ\) & \(\circ\) & \(\circ\) \\
        \hline
        Performing poorly on identifying real images from deepfake images could have negative ramifications. & \(\circ\) & \(\circ\) & \(\circ\) & \(\circ\) & \(\circ\) & \(\circ\) & \(\circ\) \\
        \hline
    \end{tabular}
\end{table}
\begin{table}[H]
    \centering
    \small % Adjust font size for compactness
    \caption{Post-Phase Trust Evaluation Assessment}
    \label{fig:post_phase_trust}
    \renewcommand{\arraystretch}{0.9} % Reduce row height
    \setlength{\tabcolsep}{2pt} % Reduce horizontal padding between columns
    \begin{tabular}{|p{3.2cm}|p{1.2cm}|p{1.2cm}|p{1.2cm}|p{1.2cm}|p{1.2cm}|p{1.2cm}|p{1.2cm}|}
        \hline
        \textbf{Statement} & \textbf{Strongly disagree} & \textbf{Disagree} & \textbf{Somewhat disagree} & \textbf{Neutral} & \textbf{Somewhat agree} & \textbf{Agree} & \textbf{Strongly agree} \\
        \hline
        If others say this is a good AI system, then it’s good for me too. & \(\circ\) & \(\circ\) & \(\circ\) & \(\circ\) & \(\circ\) & \(\circ\) & \(\circ\) \\
        \hline
        If the majority of other users trust the AI system, then I must trust it too. & \(\circ\) & \(\circ\) & \(\circ\) & \(\circ\) & \(\circ\) & \(\circ\) & \(\circ\) \\
        \hline
        If the AI system is used by a large group of people, then it is okay for me to use it too. & \(\circ\) & \(\circ\) & \(\circ\) & \(\circ\) & \(\circ\) & \(\circ\) & \(\circ\) \\
        \hline
    \end{tabular}
\end{table}

% \section{Confidence Rating By Question}
\section{Where Participants Saw Deepfakes By Group}

Each group showed a strong majority of participants reporting exposure to deepfakes through social media platforms:

\begin{table}[ht]
\centering
\caption{Where Participants Saw Deepfakes by Group}
\begin{tabular}{lcccc}
\toprule
\textbf{Source} & \textbf{Group 1 (n=97)} & \textbf{Group 2 (n=99)} & \textbf{Group 3 (n=100)} & \textbf{Group 4 (n=96)} \\
\midrule
Social Media Platforms     & 80 (82.5\%)  & 77 (77.8\%)  & 82 (82.0\%)  & 74 (77.1\%) \\
Others, Please Specify     & 7 (7.2\%)    & 10 (10.1\%)  & 9 (9.0\%)    & 11 (11.5\%) \\
News Websites              & 4 (4.1\%)    & 5 (5.1\%)    & 2 (2.0\%)    & 3 (3.1\%)   \\
Prefer Not to Answer       & 6 (7.2\%)    & 7 (7.1\%)    & 7 (7.0\%)    & 8 (8.3\%)   \\
\bottomrule
\end{tabular}
\label{tab:deepfake_exposure}
\end{table}

\section{Prior Photo Editing Experience  By Group}

There’s a diverse range of photo editing experience across groups, with many having no prior experience:

\begin{table}[ht]
\centering
\caption{Photo Editing Software Experience by Group}
\begin{tabular}{lcccc}
\toprule
\textbf{Experience Level} & \textbf{Group 1 (n=97)} & \textbf{Group 2 (n=99)} & \textbf{Group 3 (n=100)} & \textbf{Group 4 (n=96)} \\
\midrule
Yes, somewhat experienced      & 35 (36.1\%) & 42 (42.4\%) & 41 (41.0\%) & 43 (44.8\%) \\
Yes, a lot of experience       & 5 (5.1\%)   & 6 (6.1\%)   & 18 (18.0\%) & 10 (10.4\%) \\
No, never used it              & 44 (45.4\%) & 33 (33.3\%) & 25 (25.0\%) & 28 (29.2\%) \\
Yes, just started              & 13 (13.4\%) & 16 (16.2\%) & 15 (15.0\%) & 15 (15.6\%) \\
Prefer not to answer           & 0           & 2 (2.0\%)   & 1 (1.0\%)   & 0           \\
\bottomrule
\end{tabular}
\label{tab:photo_editing_experience}
\end{table}

% Social media is by far the most common source across all groups (≥77\%), with very few citing news websites or skipping the question.

% Group 3 had the highest number of participants with "a lot of experience" (18\%), compared to only 5–10\% in other groups.

% Group 1 had the highest percentage of participants with no experience (46.4\%).

% A moderate portion across all groups were somewhat experienced (37–45\%).

\clearpage

\end{document}